\documentclass[onecolumn]{jpsj2}
\usepackage{bm}
\usepackage{graphics}
\usepackage{graphicx}
\usepackage{eepic}
\usepackage{slashbox}

\renewcommand{\d}{{\rm d}}

\newcommand{\Qic}{{\bm Q}_{\rm ic}}

\title{Enhanced Spin Susceptibility toward the Charge-Ordering Transition \\
in a Two-Dimensional Extended Hubbard Model}

\author{Kazuyoshi \textsc{Yoshimi}\thanks{E-mail address: yoshimi@issp.u-tokyo.ac.jp},
Takeo \textsc{Kato} and Hideaki \textsc{Maebashi} }

\inst{Institute for Solid State Physics, University of Tokyo,
 Kashiwanoha, Kashiwa, Chiba 277-8581}

\abst{
Based on the non-skeleton diagrammatic expansion satisfying the compressibility and 
spin-susceptibility sum rules, we investigate static charge and spin responses in a 
two-dimensional extended Hubbard model with the nearest-neighbor Coulomb repulsion 
in the vicinity of its charge-ordering transition point. 
In this expansion, we can calculate approximate charge and spin response functions 
by systematic inclusion of vertex corrections, from which we obtain the uniform 
susceptibility equal to the so-called $q$-limit of the response function and  the 
second-order transition point as a divergent point in the same response function 
at some finite wave-number vector. 
It is shown that the reentrant charge-ordering transition, which has already been observed in 
the random-phase approximation (RPA), remains to take place even though the vertex 
corrections are included beyond the RPA. As a prominent effect of the vertex corrections, 
we find that the uniform spin susceptibility is enhanced due to charge fluctuations developing
toward the charge-ordering transition. 
We give a qualitative comparison of this enhanced spin susceptibility with the experimental results on the 
quasi-two-dimensional organic conductors, together with its explanation in the Landau's Fermi-liquid theory.
}

\kword
{
conserving approximations,
vertex corrections,
extended Hubbard model, 
charge ordering,
organic conductors
}

\def\runauthor{
K. \textsc{Yoshimi}, T. \textsc{Kato} and H. \textsc{Maebashi} }

\begin{document}
\maketitle

\section{Introduction}

Low-dimensional organic conductors exhibit various electronic properties such as superconductivity, 
magnetism and charge ordering. Among a variety of phase transitions appearing in organic conductors,
charge-ordering (CO) phenomena have recently attracted much attention in the context of strongly correlated 
electron systems\cite{H.Seo1}. It has been established that these CO phenomena are caused 
by long-range Coulomb interaction between conduction electrons.
Quasi-two-dimensional material $\theta$-(BEDT-TTF)$_2$RbZn$_4$ is one of typical compounds most extensively 
studied for CO. This system shows a sharp metal-insulator transition characterized by a rapid
exponential increase of the resistivity below $T_{\rm MI}=195{\rm K}$~\cite{H.Mori1}, where
a structural transition accompanying the doubling of the lattice periodicity along $c$-axis occurs.
Clear evidence of charge disproportionation in the CO phase has
been reported by NMR and Raman/infrared spectrum, whereas long-range ordering has been observed
by X-ray diffraction experiment~\cite{T.Takahashi}.

In contrast to remarkable change in the charge degree of freedom, the spin susceptibility is frequently
less sensitive to the CO transition, but exhibits non-trivial dependence on temperature. 
For example, the spin susceptibility of $\theta$-(BEDT-TTF)$_2$RbZn$_4$ has a broad peak around the
metal-insulator transition temperature $T_{\rm MI}$ without any sign of singularity. 
A paramagnetic phase is kept down to 30K at which a spin-density-wave (SDW) transition occurs, 
showing the so-called spin-gap behavior. 
This behavior of the spin susceptibility above 30K is analyzed qualitatively 
by the Bonner-Fisher curve of the Heisenberg model,
which is an effective model of localized electrons in a strong coupling region. 
Although this analysis may be justified in the insulating phase,
it is nontrivial to apply it to the metallic phase above $T_{\rm MI}$. It is also clear that
this effective spin model cannot treat the effect of charge disproportionation and its fluctuation properly. 
Oppositely, if one starts with a metallic state, and incorporates the Coulomb interaction as a perturbation,
the temperature-dependence of the spin susceptibility above $T_{\rm MI}$ is captured by correction to
temperature-independent Pauli paramagnetism. This correction is expected to be induced by development of
quantum fluctuations near the CO transition. This weak-coupling approach
has not been studied in the previous theoretical studies of the CO phenomena.

So far charge fluctuations developed in the uniform metallic phase near the CO transition points 
have mainly been studied by the random-phase approximation (RPA). In this approximation, the CO 
transition is determined by a divergence in the static charge response function at a finite wave-number vector  
since it indicates instability of the uniform metallic state. If the transition is of the second order,
the transition point thus determined agrees with the one obtained by the self-consistent Hartree approximation,
which has succeeded in obtaining various spatial patterns of CO states
in quasi-two-dimensional organic conductors\cite{H.Seo,H.Seo2}. We note that the RPA has also been 
applied to the problem of superconductivity next to the CO 
phase~\cite{A.Kobayashi1,Tanaka04, Yoshimi07, Kuroki06a,Kuroki06b}.

It is, however, emphasized that the RPA is not suitable for understanding the non-trivial temperature-dependence 
of the uniform spin susceptibility because the spin susceptibility in the RPA is not affected at all by 
enhanced charge fluctuations near the CO transition points. In order to include effect of 
this charge fluctuations, we need to study vertex corrections to the response function.
As leading vertex corrections, the Maki-Thompson (MT) type and the Aslamazov-Larkin (AL) type are 
well known. In the two-dimensional (2D) Hubbard model near the half-filling, these two types of 
vertex corrections have been investigated, respectively, to show that nested antiferromagnetic spin fluctuations 
lead to the tiny Drude weight \cite{Maebashi00,Miyake01} and the enhanced charge compressibility\cite{Miyake, Morita}.

In this paper, we investigate the vertex corrections to the charge and spin response functions in the 2D 
extended Hubbard model (EHM) at the 3/4-filling in the presence of the nearest-neighbor Coulomb repulsion.
We show that the charge fluctuations lead to an enhancement of the spin susceptibility toward the CO  
transition at which both of the MT and AL type vertex corrections give an important contribution. 
It is, here, noted that if one takes some approximation, one encounters the problem that the compressibility 
(spin-susceptibility) sum rules is not necessarily guaranteed to hold; namely, the uniform charge (spin) susceptibility 
defined by the derivative of the electron  number (magnetization) with respect to the chemical potential 
(external magnetic field) does not necessarily agree with the $q$-limit of the charge (spin) response function. 
To avoid this problem, we construct a theoretical formalism based on the {\it non-skeleton} diagrammatic conserving approximations, 
assisted by a set of the so-called Hedin's equations,\cite{Hedin65} which provide a formal scheme for iterative 
generation of a series of the {\it skeleton} diagrammatic conserving approximations\cite{Takada95}. 

In our non-skeleton diagrammatic formalism, the compressibility and spin-susceptibility sum rules are satisfied 
at each level of the approximation as in the skeleton diagrammatic conserving approximations\cite{Baym61,Baym62}, 
although our approximate response function has a rather simple form which enables us to make an actual calculation 
in the lower levels of the approximation: The 0th level of the approximation corresponds to 
the Hartree approximation for the self-energy and the RPA for the response function. 
By iterative generation applied to this 0th level,  we get the 1st level of the non-skeleton diagrammatic 
conserving approximation abbreviated by 1NSCA, in which the response function includes naturally 
the MT and AL type vertex corrections while the self-energy is described by the RPA.

This paper is organized as follows.
In \S~2, we introduce the non-skeleton diagrammatic conserving approximation to formulate 
systematic inclusion of the vertex corrections. For readers who are not interested
in details of theoretical formulation, this section may be skipped. 
In \S~3, we apply this formalism to the EHM on the 2D square lattice and then make actual 
calculations of the static charge and spin response functions in the 1NSCA. From the results 
of these response functions, we obtain the spin and charge susceptibilities in the uniform metallic 
phase near the CO transition point. 
After giving discussions including relation to the Fermi-liquid theory and comparison with the 
experiments in \S~4, a summary is given in \S~5.


\section{Formulation}

\subsection{Exact relations \label{sec:2-1}}

In this section, we consider a generic single-band model Hamiltonian ${\cal H}$ on a lattice with band dispersion
$\varepsilon_{\bm k}$ and electron-electron interaction $v_{\sigma \sigma'} ({\bm q})$ 
to formulate systematic inclusion of vertex corrections: 
\begin{align}
{\cal H} &= \sum_{{\bm k}, \sigma} (\epsilon_{\bm k}^{\mathstrut} - \mu_{\sigma}^{\mathstrut}) 
c_{{\bm k}, \sigma}^{\dag}c_{{\bm k}, \sigma}^{\mathstrut}
\nonumber
\\ 
&+ \frac{1}{2 \Omega} \sum_{\sigma, \sigma '}^{\mathstrut} \sum_{{\bm k}, {\bm k}', {\bm q}}^{\mathstrut}
v_{\sigma \sigma '}^{\mathstrut}({\bm q}) 
c_{{\bm k}+{\bm q}, \sigma}^{\dag} 
c_{{\bm k}'-{\bm q}, \sigma'}^{\dag}c_{{\bm k}', \sigma'}^{\mathstrut}
c_{{\bm k}, \sigma}^{\mathstrut}\,, 
\label{sec1; Hamiltonian}
\end{align}
where $\Omega$ is the volume of the lattice system, $c_{{\bm k} \sigma}^{\dag}$ ($c_{{\bm k} \sigma}$) is 
the creation (annihilation) operator of 
an electron with a wave number vector ${\bm k}$ and a spin $\sigma ~(= + (-)$ for $\uparrow (\downarrow)$). 
Because of usefulness for describing 
the exact relations, we have introduced $\mu_{\sigma}=\mu+\sigma h$
with $\mu$ and $h$ being a chemical potential and an external magnetic field, respectively. 
Throughout this paper, we take the limit $h \rightarrow 0$ at the end of calculation.

The charge and spin densities for the Hamiltonian ${\cal H}$ are given by $n = \sum_{\sigma} n_{\sigma}$ 
and $m = \sum_{\sigma} \sigma n_{\sigma}$, respectively. Here, $n_{\sigma}$ is the electron number density 
for a spin $\sigma$; $n_{\sigma}$ can be related to the single-particle Green's function $G_{\sigma}(k)$ by
\begin{align}
n_{\sigma}
=  \int_{k}  {\rm e}^{{\rm i}\epsilon_l \eta}G_{\sigma}(k), 
\label{density}
\end{align} 
where $\int_{k}$ denotes $(T/\Omega) \sum_{k} = (T/\Omega) \sum_{{\bm k}} \sum_l$ with 
$k$ being a combined notation of a wave number vector ${\bm k}$ and a fermionic Matsubara 
frequency ${\rm i}\epsilon_l = (2 l +1)\pi {\rm i} T$ with an integer $l$, 
and $\eta$ is a positive infinitesimal. 
We can write $G_{\sigma}(k)$ as
\begin{align}
G_{\sigma}(k) = \frac{1}{ {\rm i} \epsilon_l + \tilde{\mu}_{\sigma} - \varepsilon_{\bm k} 
- \tilde{\Sigma}_{\sigma}(k)}\,,
\label{Dyson}
\end{align}
where $\tilde{\Sigma}_{\sigma}(k)$ is the self-energy in which the Hartree contribution is subtracted
while the chemical potential is shifted by this Hartree contribution as
\begin{equation}
\tilde{\mu}_{\sigma} = \mu_{\sigma} - \sum_{\sigma'} v_{\sigma \sigma'}({\bm 0}) n_{\sigma'}.
\label{chemical potential shift}
\end{equation}

The charge and spin response functions are related to 
the density-density response function $\chi_{\sigma \sigma'}(q)$ as
\begin{subequations}
\label{def: response functions}
\begin{align}
\chi_{_{NN}}^{\mathstrut}(q)&= \frac{1}{2} \sum_{\sigma, \sigma'} \chi_{\sigma \sigma'}(q),\\
\chi_{_{S_z S_z}}^{\mathstrut}(q)&= \frac{1}{2} \sum_{\sigma, \sigma'} \sigma \sigma' \chi_{\sigma \sigma'}(q),
\end{align}
\end{subequations}
where $q$ represents a combined notation of a wave number vector 
${\bm q}$ and a bosonic Matsubara frequency ${\rm i}\omega_l = 2 l \pi {\rm i} T$ with an integer $l$; 
$\chi_{\sigma \sigma'}(q)$ can be written in terms of the one-interaction irreducible part 
$\tilde{\chi}_{\sigma \sigma'}(q)$ as
\begin{subequations} 
\label{density-density response function}
\begin{align}
\chi_{\sigma \sigma'}(q)  &= \tilde{\chi}_{\sigma \sigma'}(q) - 
\sum_{\sigma_1, \sigma_2} \tilde{\chi}_{\sigma \sigma_1}(q) v_{\sigma_1 \sigma_2}({\bm q}) \chi_{\sigma_2 \sigma'}(q),
\label{partial response function}
\\
\tilde{\chi}_{\sigma \sigma'}(q) &= - \int_{k} G_{\sigma}(k+q)G_{\sigma}(k) \Lambda_{\sigma \sigma'}(k; q).
\label{vertex corrections}
\end{align}
\end{subequations}
Here, $\Lambda_{\sigma \sigma'}(k; q)$ is the vertex function, including the vertex corrections 
to the response functions.

The uniform charge and spin susceptibilities (per spin), which are of central interest in this paper, 
can be defined by $\chi_c$ $\equiv$ $(1/2) (\partial n / \partial \mu)$ $=$ 
$(1/2) \sum_{\sigma, \sigma'} (\partial n_{\sigma} / \partial \mu_{\sigma'})$ and 
$\chi_s$ $\equiv$ $(1/2) (\partial m / \partial h)$ $=$ 
$(1/2) \sum_{\sigma, \sigma'} \sigma \sigma'  (\partial n_{\sigma} / \partial \mu_{\sigma'})$, 
respectively.
Then $\chi_c$ and $\chi_s$ should be equal to the so-called 
$q$-limit of the charge and spin response functions as
\begin{subequations}
\label{chi_identity}
\begin{align}
\chi_c &\equiv 
\frac{1}{2} 
\frac{\partial n}{\partial \mu}
= \chi_{_{NN}}^{\mathstrut}(0), 
\label{chi_charge}
\\
\chi_s&\equiv
\frac{1}{2}
\frac{\partial m}{\partial h}
= \chi_{_{S_z S_z}}^{\mathstrut}(0). 
\label{chi_spin}
\end{align}
\end{subequations}
Equations (\ref{chi_charge}) and (\ref{chi_spin}) are, respectively, called the compressibility and spin-susceptibility 
sum rules, which hold when both the response functions and the isothermal susceptibilities are exactly calculated. 
If one takes some approximation, however, these equations are not necessarily guaranteed to hold. 
In this paper, we shall take an approximation for the response function and determine the charge-ordering transition 
point such that
\begin{align}
\chi_{_{NN}}^{\mathstrut}({\bm q},0) = \infty \quad \mbox{at some finite ${\bm q}={\bm Q}^*$}.
\end{align}
In order to study the behaviors of the charge and spin susceptibilities near the transition point thus determined,
it is desirable that eqs.~(\ref{chi_identity}) should be satisfied in the approximation we shall take. 
For construction of such an approximation, it is crucial to satisfy the following $q$-limit Ward identity:
\begin{align}
\Lambda_{\sigma \sigma'}(k; 0) 
= \delta_{\sigma \sigma'} - 
\frac{\partial \tilde{\Sigma}_{\sigma}(k)}{\partial \tilde{\mu}_{\sigma'}}\ .
\label{Ward q-limit}
\end{align}
We can show that the sum rules (\ref{chi_identity}) hold automatically in arbitrary approximation
satisfying the Ward identity (\ref{Ward q-limit}) as follows:  
By eqs.~(\ref{density}), (\ref{Dyson}) and (\ref{chemical potential shift}), 
we can easily see
\begin{align}
\frac{\partial n_{\sigma}}{\partial \mu_{\sigma'}}
= & - \int_{k}  G_{\sigma}(k)^2 \sum_{\sigma_1} 
\left(  
\delta_{\sigma, \sigma_1}
- \frac{\partial \tilde{\Sigma}_{\sigma}(k)}{\partial \tilde{\mu}_{\sigma_1}} 
\right)
\frac{\partial \tilde{\mu}_{\sigma_1}}{\partial \mu_{\sigma'}}
\nonumber
\\
= &- \int_{k}  G_{\sigma}(k)^2 \sum_{\sigma_1} 
\left(  
\delta_{\sigma, \sigma_1}
- \frac{\partial \tilde{\Sigma}_{\sigma}(k)}{\partial \tilde{\mu}_{\sigma_1}} 
\right) \nonumber
\\
& \times \left(
\delta_{\sigma_1,\sigma'}
- \sum_{\sigma_2} v_{\sigma_1 \sigma_2}({\bm 0}) 
\frac{\partial n_{\sigma_2}}{\partial \mu_{\sigma'}} \right)\,.
\label{derivative_density}
\end{align}
Substituting eq.~(\ref{Ward q-limit}) into eq.~(\ref{derivative_density}) and 
using eqs.~(\ref{density-density response function}),
we can, in fact, confirm that eqs.~(\ref{chi_identity}) hold. 

Before closing this subsection, we give some comments on a set of the so-called Hedin's equations  
in the quantum many-body theory\cite{Hedin65}. 
In addition to eqs.~(\ref{density-density response function}), the following equations make up 
this set: 
\begin{subequations}
\label{Hedin's equations}
\begin{align}
\tilde{\Sigma}_{\sigma}(k) = - \sum_{\sigma'} \int_{q} G_{\sigma} (k+q) V_{\sigma \sigma'}(q) 
\Lambda_{\sigma \sigma'}(k;q)\,,
\label{excess self-energy}
\end{align}  
where $V_{\sigma \sigma'}(q)$ represents the effective electron-electron interaction given by
\begin{align}
V_{\sigma \sigma'}(q)  = v_{\sigma \sigma'}({\bm q}) 
- \sum_{\sigma_1, \sigma_2}v_{\sigma\sigma_1}({\bm q})\tilde{\chi}_{\sigma_1 \sigma_2}(q) 
V_{\sigma_2 \sigma'}(q)\,. 
\end{align}
\end{subequations}
If vertex corrections are neglected, i.e. the vertex function is approximated by 
$\Lambda_{\sigma \sigma'}(k; q) = \delta_{\sigma,\sigma'}$
in eqs.~(\ref{density-density response function})  and (\ref{Hedin's equations}), 
one gets renormalized RPA. 
It is, however, noted that neither eqs.~(\ref{chi_identity}) nor eq.~(\ref{Ward q-limit}) holds 
in the renormalized RPA. 
Generic algorithm to include vertex corrections, preserving eq.~(\ref{Ward q-limit}) and 
therefore eqs.~(\ref{chi_identity}), was proposed by Baym and Kadanoff (BK)\cite{Baym61,Baym62} 
based on the skeleton diagrammatic expansion with respect to the dressed Green's 
function. In principle, the exact self-energy can be obtained from eqs.~(\ref{density-density response function})  
and (\ref{Hedin's equations}) combined with this BK algorithm\cite{Takada95}.

In the following subsections, we introduce two approximation procedures,
both of which satisfy the Ward identity (\ref{Ward q-limit}); 
one is the well-known procedure based on the mean-field approximation in \S~2.2 and 
the other is a non-trivial approximation procedure including vertex corrections in \S~2.3. 
For details of a systematic procedure for generating a series of such approximations, see Appendix.

\subsection{Mean-field and random-phase approximations \label{sec:2-2}}

In the $0$th level of the approximation, we neglect ${\tilde \Sigma}_{\sigma} (k)$ 
in eq.~(\ref{Dyson}), i.e., ${\tilde \Sigma}_{\sigma}^{(0)} (k) = 0$,  
leading to the Hartree or mean-field approximation for the single-particle 
Green's function as
\begin{align}
G_{\sigma}^{(0)}(k) = \frac{1}{ {\rm i} \epsilon_l + \tilde{\mu}_{\sigma} - \varepsilon_{\bm k}} \,.
\end{align}
By virtue of the Ward identity (\ref{Ward q-limit}), $\Lambda^{(0)}_{\sigma,\sigma'}(k; q)$ should be 
approximated by
\begin{equation}
\Lambda^{(0)}_{\sigma,\sigma'}(k; q) = \delta_{\sigma \sigma'}.
\label{the 0th level of Lambda}
\end{equation}
Substituting eq.~(\ref{the 0th level of Lambda}) into eq.~(\ref{vertex corrections}),
we get the one-interaction irreducible response function in the $0$th level as
\begin{align}
\tilde{\chi}^{(0)}_{\sigma \sigma'}(q) \equiv \tilde{\chi}_0 (q) \delta_{\sigma,\sigma'} 
\equiv - \int_{k}G^{(0)}_{\sigma}(k+q)G^{(0)}_{\sigma}(k) \delta_{\sigma, \sigma'}\,. 
\label{the 0th level of chi}
\end{align}
Then eq.~(\ref{partial response function}) leads to the (unrenormalized) RPA for the response functions.
This approximation is, however, not appropriate to the present study of the uniform susceptibilities;
the uniform (${\bm q}=0$) and static ($\omega_l = 0$) response functions are not affected at all by 
charge fluctuations developed at a finite wave-number vector ${\bm q}={\bm Q}^*$, 
because the static response function is calculated in a closed form for a given ${\bm q}$, 
and has no mode coupling.

\begin{figure}[t]
\begin{center}
\resizebox{70mm}{!}{\includegraphics{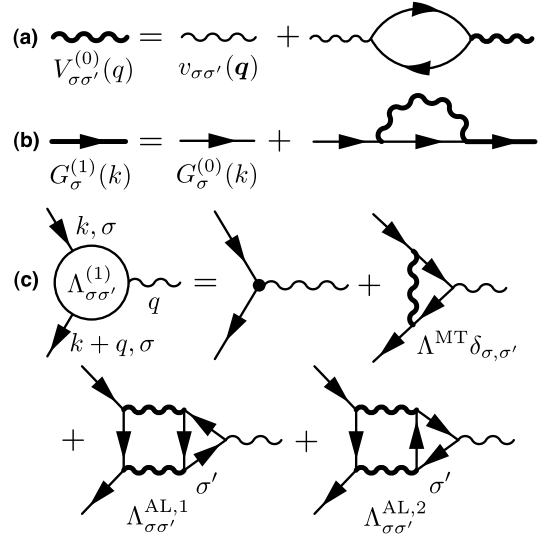}}
\end{center}
\caption{
The Feynman diagrams (a) for the effective electron-electron interaction $V^{(0)}_{\sigma \sigma'}(q)$,
(b) for the approximate single-particle Green's function $G_{\sigma}^{(1)}(k)$, and
(c) for the approximate vertex function $\Lambda_{\sigma \sigma'}^{(1)}$. 
The second term of r.h.s. in (c) corresponds to the Maki-Thompson (MT) type vertex correction, 
while the sum of the last two terms corresponds to the Aslamazov-Larkin (AL) type vertex corrections 
($\Lambda_{\sigma \sigma'}^{\rm AL}
= \Lambda_{\sigma \sigma'}^{{\rm AL},1} + \Lambda_{\sigma \sigma'}^{{\rm AL},2}$).
}
\label{fig:Diagram1.eps}
\end{figure}

\subsection{Inclusion of vertex corrections\label{sec:2-3}}

In our approximation procedure, the Ward identity (\ref{Ward q-limit}) is required to 
hold as discussed in \S~\ref{sec:2-1}.
For this purpose, it is essential to utilize the concept of the Hedin's formalism
constructed by the skeleton diagrams~\cite{Hedin65},
but we employ the non-skeleton diagrammatic formalism.
Substituting eqs.~(\ref{the 0th level of Lambda}) and (\ref{the 0th level of chi}) 
into eqs.~(\ref{Hedin's equations}), 
we get the self-energy in the $1$st level of the approximation as
\begin{equation}
\tilde{\Sigma}_{\sigma}^{(1)}(k) = - \int_{q} G_{\sigma}^{(0)} (k+q) V^{(0)}_{\sigma \sigma}(q)\,.
\end{equation}
Here $V^{(0)}_{\sigma \sigma'}(q)$ is the effective interaction in the RPA defined by
\begin{align}
V^{(0)}_{\sigma \sigma'}(q)  = v_{\sigma \sigma'}({\bm q}) 
- \sum_{\sigma_1, \sigma_2}v_{\sigma\sigma_1}({\bm q})\tilde{\chi}^{(0)}_{\sigma_1 \sigma_2}(q) 
V^{(0)}_{\sigma_2 \sigma'}(q)\,, 
\end{align}
where the diagram is shown in Fig.~\ref{fig:Diagram1.eps}(a).
The approximate Green's function is then given by
\begin{align}
[ G_{\sigma}^{(1)}(k) ]^{-1} = [ G_{\sigma}^{(0)}(k)]^{-1}- \tilde{\Sigma}_{\sigma}^{(1)}(k)\,,
\label{self-energy_ITR1} 
\end{align}
where the diagram is shown in Fig.~\ref{fig:Diagram1.eps}(b). 

The vertex function $\Lambda_{\sigma \sigma'}^{(1)}(k;q)$ carrying a momentum $q$
is constructed by replacing the internal $G^{(0)}_{\sigma'} (k')$ line in ${\tilde \Sigma}_{\sigma}^{(1)}(k)$
by $G^{(0)}_{\sigma'}(k'+q) G^{(0)}_{\sigma'} (k')$ in all the possible ways. Here, note that
all the internal momenta $k'$ are replaced so as to conserve
the wave numbers and the Matsubara frequencies at all the internal and external vertices.
The diagrams for the vertex function thus obtained are shown in
Fig.~\ref{fig:Diagram1.eps}(c). By this procedure, we can derive various types of the Ward identity,
one of which is the Ward identity (\ref{Ward q-limit}) for the $q$-limit charge vertex. 
This identity can, in fact, be proved by seeing the fact that the differential operation in eq.~(\ref{Ward q-limit}) 
is equivalent to an operation in which one first plucks out the internal 
$G_{\sigma'}^{(0)}(k')$ line in $\tilde{\Sigma}^{(1)}_{\sigma}(k)$ in all the possible ways 
and then replaces it with $G^{(0)}_{\sigma'}(p)^2$; 
this operation is nothing but the vertex insertion described above for $q=0$.

The vertex corrections can be written as the sum of the two contributions as
\begin{subequations}
\begin{align}
\Lambda_{\sigma \sigma'}^{(1)} (k; q) = \delta_{\sigma,\sigma'} + \Lambda^{\rm MT} (k; q) \delta_{\sigma,\sigma'}
+ \Lambda^{\rm AL}_{\sigma\sigma'} (k; q) \label{Lambda_ITR1} \,.
\end{align}
The Maki-Thompson (MT) type vertex correction $\Lambda^{\rm MT} (k; q)$ is written as
\begin{align}
\Lambda^{\rm MT} (k; q) = \int_{q_1} V_{\sigma \sigma}^{(0)}(q_1)
G^{(0)}_{\sigma}(k+q_1+q) G^{(0)}_{\sigma}(k+q_1) \, ,
\end{align}
while the Aslamazov-Larkin (AL) type vertex correction $\Lambda^{\rm AL}_{\sigma \sigma '} (k;q)$ is written as
\begin{align}
\Lambda^{\rm AL}_{\sigma \sigma '} (k; q) &= 
\int_{q_1}G_{\sigma}^{(0)}(k- q_1) V_{\sigma \sigma '}^{(0)}(q_1)V_{\sigma \sigma '}^{(0)}(q_1 + q) 
\nonumber\\ 
& \quad \times \left[\, \gamma_{\sigma'}(q_1; q) + c.c. \,\right]\,, 
\label{AL1}
\end{align}
where $\gamma_{\sigma}(q_1;q)$ is the fermion loop with three vertex insertions carrying momenta 
$q_1$, $q$ and $-q_1-q$ given by
\begin{align}
\gamma_{\sigma}(q_1;q) = 
- \int_{p}G_{\sigma}^{(0)}(p)G_{\sigma}^{(0)}(p + q_1)G_{\sigma}^{(0)}(p + q_1 + q)\,.
\label{gamma_ITR1}
\end{align}
\end{subequations}

\begin{figure}[t]
\begin{center}
\resizebox{70mm}{!}{\includegraphics{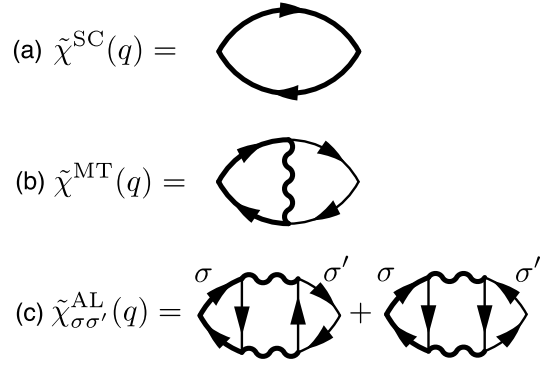}}
\end{center}
\caption{
The Feynman diagrams for the one-interaction irreducible response function
(a) without vertex corrections, (b) with the Maki-Thompson (MT) type vertex correction  
and (c) with the Aslamazov-Larkin (AL) type vertex corrections. 
The thin and thick solid lines with arrows represent the bare and dressed Green's functions, 
$G_{\sigma}^{(0)}(k)$ and $G_{\sigma}^{(1)}(k)$, respectively. 
The thick wavy lines represent the effective electron-electron interaction 
$V_{\sigma\sigma'}^{(0)}(q)$.}
\label{fig:Diagram2.eps}
\end{figure}

The one-interaction irreducible part $\tilde{\chi}_{\sigma \sigma'}^{(1)}(q)$
in the present approximation is given as
\begin{align}
\tilde{\chi}_{\sigma \sigma'}^{(1)}(q) = - \int_{k} G_{\sigma}^{(1)}(k+q)G_{\sigma}^{(1)}(k) 
\Lambda_{\sigma \sigma'}^{(1)}(k;q)\, .
\label{chi_ITR1}
\end{align}
There are three contributions in $\tilde{\chi}_{\sigma \sigma'}^{(1)}(q)$:
\begin{align}
\tilde{\chi}_{\sigma \sigma'}^{(1)}(q) = \tilde{\chi}^{\rm SC} (q) \delta_{\sigma, \sigma'}
+ \tilde{\chi}^{\rm MT} (q) \delta_{\sigma,\sigma'}
+ \tilde{\chi}^{\rm AL}_{\sigma \sigma'}(q) \, .
\end{align}
The first one corresponds to a bubble diagram which has only the self-energy correction (SC) , 
and the latter two include vertex corrections.
The Feynman diagram for each component is shown in Fig.~\ref{fig:Diagram2.eps}.
Note that these diagrams are a little different from the usual ones appearing in previous literatures;
in both $\tilde{\chi}^{\rm MT} (q)$ and $\tilde{\chi}^{\rm AL}_{\sigma \sigma'}(q)$, 
two of the Fermion lines are the {\it dressed} Green's functions 
$G_{\sigma}^{(1)}$, but all the other lines are the {\it bare} Green's functions $G_{\sigma}^{(0)}$. 
In the non-skeleton diagrammatic formalism, only this combination guarantees the compressibility 
and spin-susceptibility sum rules, eqs.~ (\ref{chi_identity}). 
In the $1$st level of the non-skeleton diagrammatic conserving approximation (1NSCA), 
the spin and charge response functions are then calculated by 
eqs.~(\ref{def: response functions}) and (\ref{density-density response function}) 
with the approximate one-interaction irreducible function $\tilde{\chi}_{\sigma \sigma'}^{(1)}(q)$. 
These response functions contain the leading 
contribution from nontrivial vertex corrections and introduce mode coupling 
between charge and spin fluctuations with different wave numbers.
Since the MT and AL type vertex corrections have a crucial role near a second-order transition 
point, we can obtain physically comprehensive results at this level of approximation. 
In the next section, we study a role of these vertex corrections near the CO transition
based on this approximation.

Finally, we mention an extension of the present approximation. We can construct
the next level of approximation by making the next self-energy from the lastly 
obtained vertex and response functions.
Performing this procedure iteratively, exact series of diagrams for all the functions are 
generated as in the skeleton-diagrammatic expansion~\cite{Takada95}, 
although it is practically difficult to sum up them.
Details of this formal extension are given in Appendix.


\section{Numerical Results and Discussions}
\subsection{Model}
In this section, we consider the single-band (extended) Hubbard model on the 2D square lattice.
In order to investigate effect of fluctuations on the response functions near the CO transition point,
we include the nearest-neighbor Coulomb interaction $V$ into the model in addition to
the on-site Hubbard interaction $U$. The Hamiltonian of the single-band EHM is given by
\begin{align}
{\cal H}_{\rm EHM} &=\sum_{{\bm k}, \sigma}(\varepsilon_{\bm k}-\mu) c_{{\bm k}\sigma}^{\dag}c_{{\bm k}\sigma}^{\mathstrut}
	+ \frac{U}{2} \sum_{i, \sigma} n_{i \sigma}n_{i \bar{\sigma}}
	+ V \sum_{\langle i j \rangle} n_i n_j ,
\end{align}
where the band dispersion is assumed to be $\varepsilon_{\bm k}=2t[\cos(k_x a)+\cos(k_y a)]$ 
with $t$ and $a$ being  the transfer integral between nearest-neighbor sites and the lattice constant, respectively. 
The chemical potential $\mu$ is determined to fix $n=\sum_{\sigma}n_{\sigma}=3/2$ by using Eq.~(\ref{density}) 
and $\bar{\sigma}$ denotes the spin with a direction opposite to $\sigma$. 
In the following, we set $t=1$, $a=1$ and $k_{\rm B}=1$. 

The Hamiltonian ${\cal H}_{\rm EHM}$ corresponds to a special case of the generic single-band model Hamiltonian 
(\ref{sec1; Hamiltonian}) with the bare Coulomb interactions given by
\begin{subequations}
\begin{align}
v_{\sigma \sigma}({\bm q})&=V({\bm q}), \\
v_{\sigma \bar{\sigma}}({\bm q})&=U+V({\bm q}),
\end{align}
\end{subequations}
where $V({\bm q})=2V(\cos q_x+\cos q_y)$. 
For the charge and spin channels, the bare interactions are represented by 
$v_c ({\bm q}) \equiv v_{\sigma \sigma}({\bm q})+v_{\sigma \bar{\sigma}}({\bm q})=U+2V({\bm q})$ 
and $v_s ({\bm q}) \equiv v_{\sigma \sigma}({\bm q})-v_{\sigma \bar{\sigma}}({\bm q})= - U$, respectively.
Then the charge and spin response functions can be written in the form 
\begin{subequations} 
\label{sus_1}
\begin{align}
\chi_{_{NN}} (q)&= 
\frac{\tilde{\chi}_{_{NN}} (q)}{1+v_c({\bm q}) \tilde{\chi}_{_{NN}} (q)}, 
\label{charge_sus_1}\\
\chi_{_{S_z S_z}} (q)&= 
\frac{\tilde{\chi}_{_{S_z S_z}} (q)}{1+v_s({\bm q}) \tilde{\chi}_{_{S_z S_z}} (q)}, 
\label{spin_sus_1}
\end{align} 
\end{subequations}
where  
$\tilde{\chi}_{_{NN}} (q) \equiv \tilde{\chi}_{\sigma \sigma}(q)+\tilde{\chi}_{\sigma \bar{\sigma}}(q)$ and 
$\tilde{\chi}_{_{S_z S_z}} (q) \equiv \tilde{\chi}_{\sigma \sigma}(q)-\tilde{\chi}_{\sigma \bar{\sigma}}(q)$.

In the RPA described in \S~2.2, $\tilde{\chi}_{_{NN}} (q)$ and $\tilde{\chi}_{_{S_z S_z}} (q)$ are approximated  
by $\tilde{\chi}_{_{NN}}^{(0)} (q) = \tilde{\chi}_{_{S_z S_z}}^{(0)} (q) = \tilde{\chi}_0(q)$, 
where $\tilde{\chi}_0(q)$ is defined by eq.~(\ref{the 0th level of chi}). 
In the 1st level of the non-skeleton diagrammatic conserving approximation (1NSCA) described in \S~2.3, 
they are approximated by
\begin{subequations} 
\label{chi_1NSCA}
\begin{align}
\tilde{\chi}_{_{NN}}^{(1)} (q)&= 
\tilde{\chi}^{\rm SC}(q)  + \tilde{\chi}^{\rm MT}(q) 
+ \tilde{\chi}^{\rm AL}_{\sigma \sigma}(q) + \tilde{\chi}^{\rm AL}_{\sigma \bar{\sigma}}(q), 
\\
\tilde{\chi}_{_{S_z S_z}}^{(1)} (q)&= 
\tilde{\chi}^{\rm SC}(q)  + \tilde{\chi}^{\rm MT}(q) 
+ \tilde{\chi}^{\rm AL}_{\sigma \sigma}(q) - \tilde{\chi}^{\rm AL}_{\sigma \bar{\sigma}}(q),
\end{align} 
\end{subequations}
where $\tilde{\chi}^{\rm SC}(q)$ is a contribution associated with a bubble diagram (Fig.~\ref{fig:Diagram2.eps}(a))
which has only the self-energy correction (SC)  
while $\tilde{\chi}^{\rm MT}(q)$ and $\tilde{\chi}^{\rm AL}_{\sigma \sigma'}(q)$ include 
the Maki-Thompson (MT) and Aslamazov-Larkin (AL) type vertex corrections 
(Figs.~\ref{fig:Diagram2.eps}(b) and \ref{fig:Diagram2.eps}(c)), respectively. 
Note that the difference between the one-interaction irreducible response functions $\tilde{\chi}_{_{NN}}^{(1)} (q)$ 
and $\tilde{\chi}_{_{S_z S_z}}^{(1)} (q)$ comes only from the AL contribution.

In order to calculate the convolution form, we use the Fast-Fourier-Transform (FFT) algorithm. 
The first Brillouin zone is divided into $64 \times 64$ meshes. The frequency sum is terminated 
at $\epsilon_{c}$ whose value is about 40 times as large as the band-width $W = 8t$ for $T=0.1$. 
In this section, we first determine the phase diagram on the $(V, T)$ plane. Next, we calculate the 
$V$-dependence for the uniform spin and charge susceptibilities. 
In the following, we fix $U=3$, at which a SDW instability is not observed for $T \ge 0.05$.

\begin{figure}[t]
\begin{center}
     \resizebox{85mm}{!}{\includegraphics{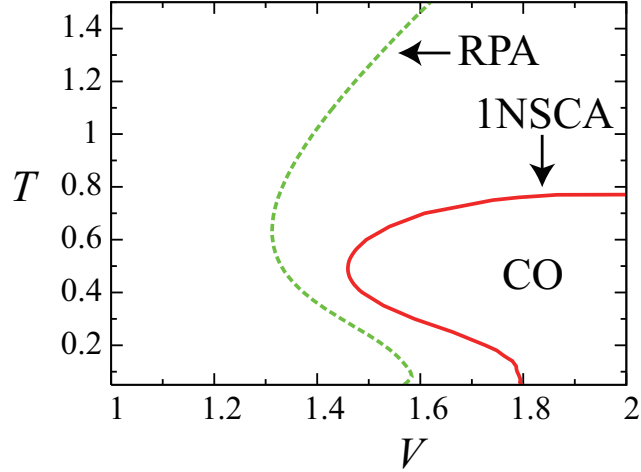}}
  \end{center}
  \caption{Phase diagram on the ($V$, $T$) plane at $U=3$. 
The dashed and solid curves represent the CO transition curves in the RPA and in the 1NSCA, respectively.}
    \label{phase_U3.eps}
\end{figure}

\begin{figure}[t]
\begin{center}
     \resizebox{85mm}{!}{\includegraphics{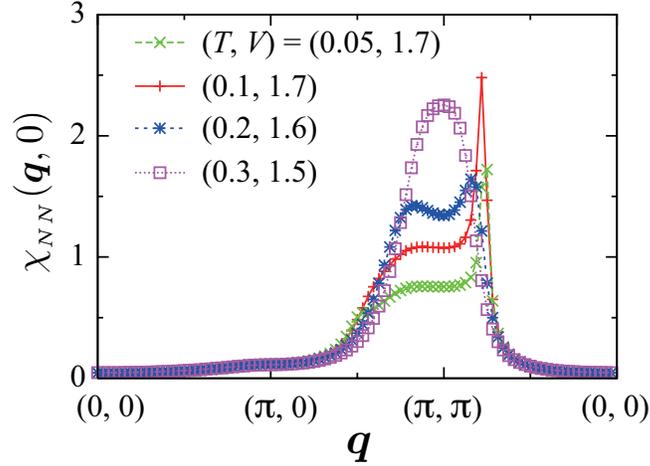}}
  \end{center}
  \caption{Static charge response function in the 1NSCA for $(T, V) = (0.05, 1.7), (0.1, 1.7), (0.2, 1.6)$ and 
$(0.3, 1.5)$ at $U=3$. 
The peak position of the response function continuously shifts from ${\bm Q}_{\rm cb}=(\pi, \pi)$ to 
$\Qic \approx (3\pi /4, 3\pi/4)$ for $T < 0.3$.}
    \label{peak_chiC.eps}
\end{figure}

\subsection{Phase diagram}

The phase diagram on the $(V, T)$ plane at $U=3$ is shown in Fig.~\ref{phase_U3.eps}, where
the CO transition curves are determined by the divergence of the static charge response function, 
i.e., by $1+v_c({\bm q}) \tilde{\chi}_{_{NN}}({\bm q},0) = 0$ in the RPA (the dashed curve) and in the 
1NSCA (the solid curve). 
Since the region of the CO state in the 1NSCA is smaller than that in the RPA, 
we can understand that the CO transition is suppressed by the charge fluctuations. 
It is, however, noted that the overall shapes of the transition curves in these approximations 
are similar to each other. 
In particular, 
the reentrant transition (metallic $\rightarrow$ CO $\rightarrow$ metallic with decreasing temperature) 
are observed both in the RPA \cite{A.Kobayashi1,  J.Merino1} and in the 1NSCA;
namely, the reentrant CO behavior remains even if the vertex corrections are taken into account\cite{footnote2}. 

In Fig.~\ref{peak_chiC.eps}, we show the static charge response function in the 1NSCA for 
the uniform metallic state near the CO transition curve in the phase diagram.
At high temperatures, the ${\bm q}$-dependence of $\tilde{\chi}_{_{NN}}({\bm q},0)$ 
does not depend on ${\bm q}$ very much, resulting in checkerboard-type charge ordering at 
${\bm q} = {\bm Q}_{\rm cb} \equiv (\pi, \pi)$ at which $v_c ({\bm q})$ takes a minimum value.
With decreasing temperature, the existence of the Fermi surface causes 
the strong ${\bm q}$-dependence of $\tilde{\chi}_{_{NN}}({\bm q})$, 
so that the peak position of the static charge response function changes
from ${\bm q}={\bm Q}_{\rm cb}$ to ${\bm q}={\bm Q}_{\rm ic} \approx (3\pi /4, 3\pi /4)$
for $T \lesssim 0.2$, leading to an incommensurate CO instability at ${\bm q} = \Qic$. 
The reentrant transition observed in Fig.~\ref{phase_U3.eps} is due to nonmonotonic
temperature-dependence of $\chi_{_{NN}}({\bm Q}_{\rm cb},0)$ with a fixed $V$;
namely, as temperature increses, $\chi_{_{NN}}({\bm Q}_{\rm cb},0)$ once increases
until $T \lesssim 0.5$ but decreases at high temperatures.
This unusual increase of $\chi_{_{NN}}({\bm Q}_{\rm cb},0)$ for low temperatures originates from
broadening of the incommensurate peak of $\chi_{_{NN}}({\bm q},0)$ at ${\bm q} = \Qic$
\begin{figure}[b]
\begin{center}
     \resizebox{85mm}{!}{\includegraphics{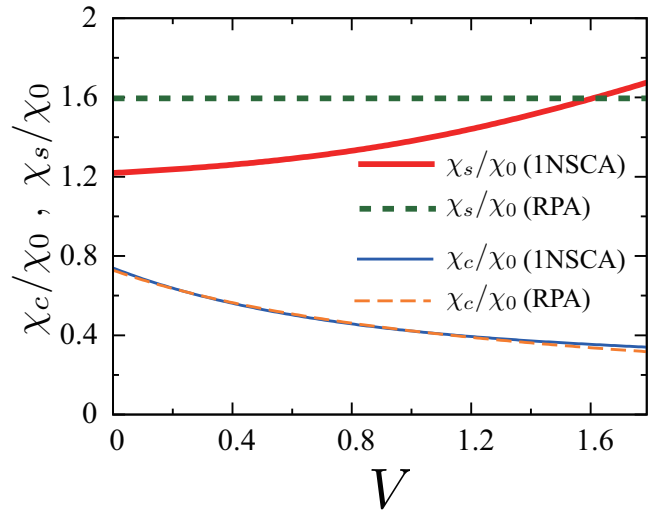}}
  \end{center}
  \caption{Uniform spin and charge susceptibilities $\chi_s$ and $\chi_c$ divided by 
the non-interacting susceptibility $\chi_0$ versus the nearest-neighbor Coulomb repulsion $V$ 
for $T=0.1$ and $U=3$. $\chi_s/\chi_0$ and $\chi_c/\chi_0$ are plotted by thick and thin solid 
curves, respectively, in the 1NSCA and by thick and thin dashed curves in the RPA.     
Note that the incommensurate CO transition occurs at $V = 1.788$ in the 1NSCA.}
    \label{static_sus_T0.1_U_3.eps}
\end{figure}

\subsection{Uniform susceptibilities}

Let us move on the main result of this paper. In our approximations, the compressibility and 
spin-susceptibility sum rules (\ref{chi_identity}) are satisfied, so that the uniform charge and spin susceptibilities 
are equal to the $q$-limit of the response functions as $\chi_c = \chi_{_{NN}} (0)$ and 
$\chi_s = \chi_{_{S_z S_z}} (0)$, respectively. 
In Fig.~\ref{static_sus_T0.1_U_3.eps}, 
we depict the results for $\chi_s/\chi_0$ (thick solid curve) and $\chi_c/\chi_0$ (thin solid curve) 
in the 1NSCA for $T=0.1$ and $U=3$ where the non-interacting susceptibility is denoted by 
$\chi_0 \equiv \tilde{\chi}_0(0)$. 
In the same figure, we also represent the results for $\chi_s/\chi_0$ 
(thick dashed line) and $\chi_c/\chi_0$ (thin dashed curve) in the RPA for comparison. 
Note that the CO transition occurs at $V=1.788$ in the 1NSCA for $T=0.1$ and $U=3$. 

From Fig.~\ref{static_sus_T0.1_U_3.eps}, 
we find that the spin susceptibility $\chi_s$ calculated by the 1NSCA 
is enhanced toward the CO transition point of $V=1.788$.
Since $\chi_s$ is independent of $V$ in the RPA, 
we can conclude that this enhancement is
caused by the vertex corrections. 
On the other hand, the charge susceptibility $\chi_c$ decreases with the increase of $V$ 
both in the RPA and in the 1NSCA;   
the effect of vertex corrections on $\chi_c$ is small as indicated 
by the fact that there is only a slight difference between the results of 
$\chi_c$ in these two approximations. 

For detailed explanations on these $V$-dependences of $\chi_c$ and $\chi_s$, 
we rewrite the $q$-limit of the self-energy, MT and AL contributions in eq~(\ref{chi_1NSCA})  
as $\tilde{\chi}_c^{\rm SC} = \tilde{\chi}_s^{\rm SC} \equiv \tilde{\chi}^{\rm SC}(0)$, 
$\tilde{\chi}_c^{\rm MT} =\tilde{\chi}_s^{\rm MT} \equiv \tilde{\chi}^{\rm MT}(0)$, 
$\tilde{\chi}_c^{\rm AL} \equiv \tilde{\chi}_{\sigma \sigma}^{\rm AL}(0) + \tilde{\chi}_{\sigma \bar{\sigma}}^{\rm AL}(0) $ 
and $\tilde{\chi}_s^{\rm AL} \equiv \tilde{\chi}_{\sigma \sigma}^{\rm AL}(0) - \tilde{\chi}_{\sigma \bar{\sigma}}^{\rm AL}(0)$ 
for simplicity. 
In the following subsections, we will investigate each of these contributions to the spin and charge susceptibilities 
individually.  

\subsubsection{Spin susceptibility}

\begin{figure}[t]
\begin{center}
     \resizebox{85mm}{!}{\includegraphics{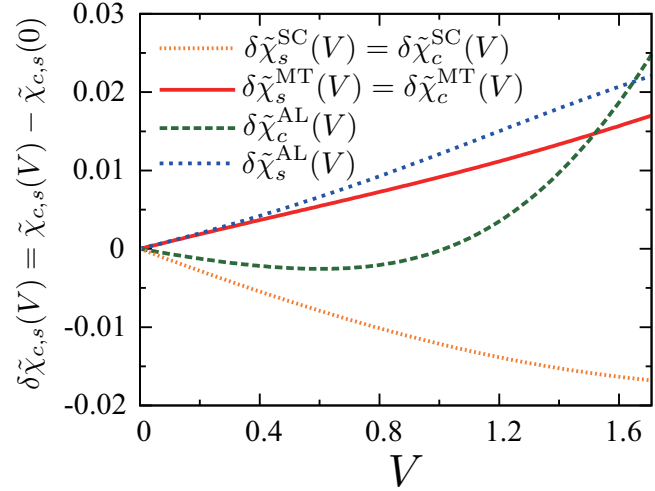}}
  \end{center}
  \caption{The $V$-dependences of $\tilde{\chi}_s^{\rm SC}$ [$=$ $\tilde{\chi}_c^{\rm SC}$], 
$\tilde{\chi}_s^{\rm MT}$ [$=$ $\tilde{\chi}_c^{\rm MT}$], $\tilde{\chi}_s^{\rm AL}$ and 
$\tilde{\chi}_c^{\rm AL}$ shifted by their values at $V = 0$. 
The temperature and the on-site Coulomb interaction are fixed 
as $T=0.1$ and $U=3$, respectively, where the CO transition occurs at $V = 1.788$.}
    \label{chiC_k0_Diagram.eps}
\end{figure}

The uniform spin susceptibility is calculated in the 1NSCA as
\begin{align}
\chi_s &= \frac{1}{\tilde{\chi}_{_{S_z S_z}}^{(1)}(0)^{-1} - U},
\label{eq:static_chi_s}
\end{align}
where $\tilde{\chi}_{_{S_z S_z}}^{(1)}(0) = \tilde{\chi}_s^{\rm SC} + \tilde{\chi}_s^{\rm MT} + \tilde{\chi}_s^{\rm AL}$. 
From eq.~(\ref{eq:static_chi_s}), it is clear that the $V$-dependence of
$\chi_s$ comes only from $\tilde{\chi}_{_{S_z S_z}}^{(1)}(0)$. Therefore, for a positive $U$,
the enhancement of $\chi_s$ observed in Fig.~\ref{static_sus_T0.1_U_3.eps} is caused by the increase 
of $\tilde{\chi}_{_{S_z S_z}}^{(1)}(0)$.
In Fig.~\ref{chiC_k0_Diagram.eps}, the $V$-dependence of 
$\tilde{\chi}_s^{\rm SC}$, $\tilde{\chi}_s^{\rm MT}$ and $\tilde{\chi}_s^{\rm AL}$
are shown by subtracting their values at $V=0$, respectively. 
It is found that, as $V$ increases, $\tilde{\chi}_s^{\rm AL}$ and $\tilde{\chi}_s^{\rm MT}$ increase, 
while $\tilde{\chi}_s^{\rm SC}$ decreases.

\begin{figure}[t]
\begin{center}
\resizebox{85mm}{!}{\includegraphics{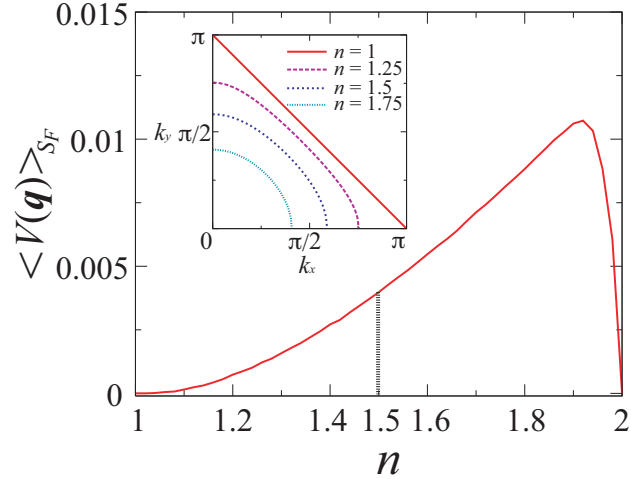}}
\end{center}
\caption{
 The $n$-dependence of the average of  
the nearest-neighbor Coulomb interaction taken over the Fermi surface: 
The inset shows the $n$-dependence of the Fermi surface for the non-interacting system.}
 \label{FS_ndep.eps}
\end{figure}

First, we discuss the behavior of $\tilde{\chi}_s^{\rm SC}$ and $\tilde{\chi}_s^{\rm MT}$.
Up to the leading correction with respect to $U$ and $V$, they can be written separately as
\begin{subequations}
\begin{align}
\tilde{\chi}^{\rm SC}_{s} & = \tilde{\chi}_0 + \tilde{\chi}^{\rm SC}_{U}+\tilde{\chi}^{\rm SC}_{V}, \\
\tilde{\chi}^{\rm MT}_{s} & = \tilde{\chi}^{\rm MT}_{U}+\tilde{\chi}^{\rm MT}_{V}.
\end{align}
\end{subequations}
Since we are interested in the $V$-dependence of the spin susceptibility, we only focus on 
$\tilde{\chi}^{\rm SC}_{V}$ and $\tilde{\chi}^{\rm MT}_{V}$, which are proportional to $V$. 
At low temperatures, these corrections are expressed by the integral of the interaction on the
Fermi surface $S_F$ as~\cite{footnote4}
\begin{subequations}
\begin{align}
\tilde{\chi}^{\rm SC}_{V} &= - \tilde{\chi}_{V}^{\rm MT} \label{SC_F2} \\
\tilde{\chi}^{\rm MT}_{V} &= \int \d {\bm k} \d {\bm k'} V({\bm k} -{\bm k'})\delta (\varepsilon_{\bm k}-\mu)\delta (\varepsilon_{{\bm k}'}-\mu)\nonumber\\
&\equiv \langle V({\bm q}) \rangle_{S_F}. \label{MT_F2}
\end{align}
\end{subequations}
The sign of $\langle V({\bm q}) \rangle$ depends on both $V({\bm q})$ and the shape of the Fermi surface.
In the present simple model with nearest-neighbor Coulomb interaction on the square lattice, the potential
$V({\bm q}) = 2V(\cos q_x +\cos q_y)$ is positive for small momentum transfer around ${\bm q} \sim 0$,
and negative for large momentum transfer around ${\bm q} \sim (\pi, \pi)$. We show
$\langle V({\bm q})\rangle_{S_F}$ for $1 \leq n \leq 2$ in Fig.~\ref{FS_ndep.eps},
together with the change of the Fermi surface in the inset. 
At the half filling ($n=1$), $\langle V({\bm q}) \rangle$ becomes zero because positive contribution
of small momentum transfer is canceled by negative one of large momentum transfer. As the filling increases,
the Fermi surface becomes smaller, and therefore negative component is reduced.
As a result, $\langle V({\bm q}) \rangle$ becomes positive, and increases until $n \simeq 1.9$.
Finally, $\langle V({\bm q}) \rangle$ rapidly decreases toward $n=2$ where the Fermi surface disappears. 
We note that the above discussion is also applicable for $0 \leq n \leq 1$ by replacing $n$ with $2-n$. 
Thus, at least in the leading correction with respect to $V$, the MT contribution $\tilde{\chi}^{\rm MT}_{V}$ is
positive for the $3/4$-filling ($n=3/2$), leading to the enhancement in the spin susceptibility. 
It is, however, noted that by eq.~(\ref{SC_F2}), this enhancement due to the MT contribution is 
canceled by the self-energy contribution $\tilde{\chi}^{\rm SC}_{V}$ for small $V$. 
These features of  $\tilde{\chi}^{\rm SC}_{V}$ and $\tilde{\chi}^{\rm MT}_{V}$ are consistent with 
the results shown in Fig.~\ref{chiC_k0_Diagram.eps}.

Next, we discuss $\tilde{\chi}_s^{\rm AL}$ in terms of the effective interaction
$V_{\sigma \sigma'}^{\rm RPA}({\mib q}) \equiv V_{\sigma \sigma'}^{(0)}({\mib q},0)$. 
Since the MT contribution is almost canceled by the self-energy one even near the CO  
transition point of $V = 1.788$ as seen in Fig.~\ref{chiC_k0_Diagram.eps},
the net enhancement of the spin susceptibility toward the CO transition results mainly 
from $\tilde{\chi}_s^{\rm AL}$. 
In the AL type diagrams shown in Fig.~2(c), 
the two of the effective interactions (the thick wavy lines) carry the same momenta 
for the uniform ($q = 0$) susceptibility and their static components around the ordering vector 
${\mib Q}^*$ are dominant near the CO transition point;  
on the other hand, the Green's functions themselves (the thick or thin solid lines) 
are less sensitive to the increase of charge fluctuations and give positive contribution to 
$\tilde{\chi}_s^{\rm AL}$. 
We can then make a rough estimation of $\tilde{\chi}_{s} ^{\rm AL}$ as
\begin{align}
\tilde{\chi}_{s}^{\rm AL} & \propto \frac{1}{2}\sum_{\sigma, \sigma'} \sigma \sigma' 
V_{\sigma \sigma'}^{\rm RPA} ({\bm Q^*})^2
\nonumber
\\
& = V_{c}^{\rm RPA}({\bm Q}^*)V_{s}^{\rm RPA}({\bm Q}^*) ,
\label{AL_Veff}
\end{align}
where $V_{c}^{\rm RPA}({\bm q})$ and $V_{s}^{\rm RPA}({\bm q})$ are given by
\begin{subequations}
\label{Veff_RPA}
\begin{align}
&V_c^{\rm RPA}({\bm q})=\frac{v_c({\bm q})}{1+v_c({\bm q}) \tilde{\chi}_0 ({\bm q}, 0)},\\
&V_s^{\rm RPA}({\bm q})=\frac{v_s({\bm q})}{1+v_s({\bm q}) \tilde{\chi}_0 ({\bm q}, 0)}.
\end{align}
\end{subequations}  
It is noted that in the paramagnetic phase without charge ordering,
the denominators in eqs.~(\ref{Veff_RPA}), i.e.,
$1+v_c({\bm Q}^*) \tilde{\chi}_0 ({\bm Q}^*, 0)$ and
$1+v_s({\bm Q}^*) \tilde{\chi}_0 ({\bm Q}^*, 0)$ are both positive.
Because charge fluctuations are developed only for the wave-numbers satisfying
$v_c({\bm Q}^*) = U + 2 V({\bm Q}^*) < 0$,
we can then see that amplitude of $V_{c}^{\rm RPA}({\bm Q}^*)$ is developed toward
the CO transition point keeping its sign negative. On the other hand,
$V_{s}^{\rm RPA}({\bm Q}^*)$ is always negative because $v_s({\bm Q}^*) = - U < 0$ for a positive $U$.
Therefore, eq.~(\ref{AL_Veff}) indicates that $\tilde{\chi}_{s}^{\rm AL}$ increases toward the
CO transition; this behavior is consistent with the result in Fig.~\ref{chiC_k0_Diagram.eps}.

In conclusion, charge fluctuations developed near the CO transition point 
make the uniform spin susceptibility increase through the vertex corrections; 
the behavior of $\chi_s$ in Fig.~\ref{static_sus_T0.1_U_3.eps} can, in fact, be understood 
qualitatively by the increase of $\tilde{\chi}_{s}^{\rm AL}$.  

\subsubsection{Charge susceptibility}

The uniform charge susceptibility is calculated in the 1NSCA as
\begin{align}
\chi_c = \frac{1}{\tilde{\chi}_{_{NN}}^{(1)}(0)^{-1} + U + 8V},
\label{eq:charge_sus}
\end{align}
where $\tilde{\chi}_{_{NN}}^{(1)}(0) = \tilde{\chi}_c^{\rm SC} + \tilde{\chi}_c^{\rm MT} + \tilde{\chi}_c^{\rm AL}$. 
The overall $V$-dependence of $\chi_c$ is dominated by $8V$, i.e., the last term in 
the denominator in eq.~(\ref{eq:charge_sus}),  so that $\chi_c$ decreases against the 
increase of $V$ as we have seen in Fig.~\ref{static_sus_T0.1_U_3.eps}. 
It is, however, interesting to consider the effect of charge fluctuations on the irreducible 
charge susceptibility $\tilde{\chi}_{_{NN}}^{(1)}(0)$. 
This consideration may suggest that $\chi_c$ increases eventually towards the CO transition 
if the higher-order vertex corrections are properly taken into account beyond the 1NSCA. 
 
Noting that $\tilde{\chi}_c^{\rm SC} = \tilde{\chi}_s^{\rm SC}$ and 
$\tilde{\chi}_c^{\rm MT} = \tilde{\chi}_s^{\rm MT}$, 
we see that the self-energy and MT contributions almost cancel each other 
in the charge susceptibility as well as in the spin susceptibility discussed in the previous subsection;   
an important difference between the irreducible charge and spin susceptibilities is then ascribed to 
the AL contribution.  
By using the same approximation in the previous subsection, 
we can make a rough estimation of $\tilde{\chi}_{c}^{\rm AL}$ as
\begin{align} 
\tilde{\chi}_c^{\rm AL} & \propto
\frac{1}{2}\sum_{\sigma, \sigma'} V_{\sigma \sigma'}^{\rm RPA} ({\bm Q^*})^2 
\nonumber
\\
& = \left[V_{c}^{\rm RPA}({\bm Q}^*)\right] ^2+\left[V_{s}^{\rm RPA}({\bm Q}^*)\right] ^2. 
\label{AL_Vc_eff}
\end{align}
In contrast to $\tilde{\chi}_s^{\rm AL}$ dependent linearly on $V_{c}^{\rm RPA}({\bm Q}^*)$, 
eq.~(\ref{AL_Vc_eff}) indicates that 
$\tilde{\chi}_{c}^{\rm AL}$ increases toward the CO transition 
in proportion to the square of $V_{c}^{\rm RPA}({\bm Q}^*)$.
Therefore, $\chi_c$ is expected to increase when $\tilde{\chi}_{c}^{\rm AL}$ changes considerably 
to overcome the increase of $8V$ in eq.~(\ref{eq:charge_sus}). In order to study this possibility in details, 
charge fluctuations should be fully treated beyond the present approximation. 
This problem will be studied elsewhere.~\cite{YKM2}

\section{Discussions}

\subsection{Relation to the Fermi-liquid theory}
In this paper, the spin and charge susceptibilities are studied up to
the leading vertex corrections with respect to charge fluctuations.
We expect that the present result gives a correct behavior of the susceptibilities
toward the CO transition until the charge fluctuations are not developed so much. 
Just near the transition, however, the higher-order vertex corrections may become relevant due to
strong charge fluctuations. Although the treatment of the higher-order corrections is 
beyond the scope of this paper, it is feasible to summarize its effect in terms of the Fermi-liquid theory,
which is a natural description of the metallic phase. 

Following a standard microscopic derivation of the isotropic Fermi liquid,~\cite{AGD} 
the uniform susceptibilities may be written as
\begin{subequations}
\begin{align}
\chi_{c} &= \frac{ {\tilde \chi}_{_{NN}} (0)}{1+v_c({\bm 0}){\tilde \chi}_{_{NN}} (0)}
=\frac{\rho}{1+\rho f_0^s}, 
\label{FL_chic}
\\
\chi_{s} &= \frac{ {\tilde \chi}_{_{S_z S_z}} (0)}{1+v_s({\bm 0}){\tilde \chi}_{_{S_z S_z}} (0)}
= \frac{\rho}{1+\rho f_0^a}, 
\label{FL_chis}
\end{align}
\end{subequations}
where $f_0^s$ and $f_0^a$ are the charge and spin channels of the Landau's quasi-particle interaction, respectively, 
and $\rho$ is the renormalized density of states at the Fermi level in proportion to the effective mass of the quasi-particle. 
For comparison with the results in the last section, it is useful to introduce the "screened" quasi-particle interaction 
in the mean field by 
$\tilde{f}_0^s \equiv f_0^s - v_c({\bm 0}) = f_0^s - U - 8V$ and 
$\tilde{f}_0^a \equiv f_0^a - v_s({\bm 0}) = f_0^a + U$. 
We can then write the irreducible charge and spin susceptibilities as
\begin{subequations}
\label{FL_chi_MF}
\begin{gather}
\tilde{\chi}_{_{NN}}(0) = \frac{ \rho}{1+\rho \tilde{f}_0^s}, \label{FL_chic_MF}\\
\tilde{\chi}_{_{S_z S_z}}(0) = \frac{ \rho}{1+\rho \tilde{f}_0^a}. \label{FL_chis_MF}
\end{gather}
\end{subequations}
For $|\rho \tilde{f}_0^s| \ll 1$ and $|\rho \tilde{f}_0^a| \ll 1$, in particular, 
we can expand these irreducible susceptibilities as follows:
\begin{subequations}
\label{expansion_FLT}
\begin{align}
\tilde{\chi}_{_{NN}}(0) &\approx \rho(1-\rho \tilde{f}_0^s)
=\rho[1-\rho (\tilde{f}_0^{\rm MT}+ \tilde{f}_0^{s,\rm AL})],
\label{expansion_FLT_a}
\\
\tilde{\chi}_{_{S_z S_z}}(0) &\approx \rho(1-\rho \tilde{f}_0^a)
=\rho [1-\rho ( \tilde{f}_0^{\rm MT}+ \tilde{f}_0^{a,\rm AL})].
\label{expansion_FLT_b}
\end{align}
\end{subequations}
In the above expansion, 
noting that the self-energy correction is included in $\rho$, 
we have decomposed $\tilde{f}_0^s$ and $\tilde{f}_0^a$ into
contributions from the MT and AL type vertex corrections as 
$\tilde{f}_0^s = \tilde{f}_0 ^{\rm MT} + \tilde{f}_0 ^{s ,{\rm AL}}$ 
and $\tilde{f}_0^a = \tilde{f}_0 ^{\rm MT} + \tilde{f}_0 ^{a ,{\rm AL}}$, 
respectively. 

\begin{table}
	\begin{tabular}{|| ll | c | c | c ||}
	\hline
	&& $\tilde{f}_0^{\rm MT}$ & $\tilde{f}_0^{s, {\rm AL}}$ & $\tilde{f}_0^{a, {\rm AL}}$\\
	\hline
	(a)~charge:&$v_c ({\bm Q}^*) < 0$ & $-$ & $-$ & $-$ \\
	\hline
	(b)~spin:&$v_c ({\bm Q}^*) > 0$ & $+$ & $-$ & $+$ \\
	\hline
	\end{tabular}
\caption{Trends toward an increase ($+$) and a decrease ($-$) in the "screened" quasi-particle interaction 
(a) for enhanced charge fluctuations, and (b) for enhanced spin fluctuations, based on the single-band EHM.
Note that $v_c ({\bm Q}^*)$ is necessarily negative in (a), but  $v_c ({\bm Q}^*)$ is assumed to be positive in (b), 
with ${\bm Q}^*$ being the ordering vector at which either charge or spin fluctuations are developed.}
\label{Fig_vertex.eps}
\end{table}

In Table~\ref{Fig_vertex.eps}(a),
we summarize the effects of the leading vertex corrections on the uniform susceptibilities in terms of
$\tilde{f}_0^{\rm MT}$, $\tilde{f}_0^{s, {\rm AL}}$ and $\tilde{f}_0^{a, {\rm AL}}$
by comparing the results in \S~3.3 to eqs.~(\ref{expansion_FLT}) for enhanced charge fluctuations.
In Table~\ref{Fig_vertex.eps}(b), in order to emphasize a characteristic change due to these charge fluctuations,
we also show a change in the same quantities when not charge but spin fluctuations are developed
near a SDW transition point.
In the latter case, the Fermi-liquid corrections have already been discussed in details,
based on the Hubbard model.\cite{Fuseya}
Note that in the present calculation for the EHM, the spin and charge fluctuations can be controlled
simply by changing $U$ and $V$, respectively.

From Table~\ref{Fig_vertex.eps}, we see that both charge and spin fluctuations make  
$\tilde{f}_0^{s, {\rm AL}}$ decrease in eq.~(\ref{expansion_FLT_a}). 
This is clear from the fact that $\tilde{\chi}_c^{{\rm AL}}$ increases with development 
of either charge or spin fluctuations as in eq.~(\ref{AL_Vc_eff}). 
The important point is that $\tilde{f}_0^{\rm MT}$ and $\tilde{f}_0^{a, {\rm AL}}$ in 
eq.~(\ref{expansion_FLT_b}) tend to decrease for enhanced charge fluctuations, but to increase for 
enhanced spin fluctuations; in other words, the charge and spin fluctuations  
provide attractive and repulsive parts in the spin channel of the quasi-particle interaction, respectively. 
The decrease in $\tilde{f}_0^{\rm MT}$ results from the exchange effect due to the nearest-neighbor 
Coulomb interaction $V$, which is absent in the Hubbard model, as we have seen in eq.~(\ref{MT_F2}). 
But this decrease tends to be canceled by the decrease in $\rho$ originating from the self-energy correction, 
so that the increase of $\chi_s$ is controlled by $\tilde{f}_0^{a, {\rm AL}}$.

For $\tilde{f}_0^{a, {\rm AL}}$, eq.~(\ref{AL_Veff}) is available not only when charge fluctuations 
are developed but also when spin fluctuations are developed.  
Noting that charge (spin) channel of the effective interaction is attractive near the CO (SDW) transition point, 
it is not hard to see from this equation that the sign of $\tilde{f}_0^{a, {\rm AL}}$ coincides with that of the spin (charge) channel 
of the interaction. 
For enhanced charge fluctuations ($V_{c}^{\rm RPA}({\bm Q}^*) < 0$), 
$V_{s}^{\rm RPA}({\bm Q}^*)$ is necessarily negative because of $v_s ({\bm Q}^*) = - U  < 0$, 
so that $\tilde{f}_0^{a, {\rm AL}}$ decreases. 
For enhanced spin fluctuations ($V_{s}^{\rm RPA}({\bm Q}^*) < 0$), on the other hand, 
$V_{c}^{\rm RPA}({\bm Q}^*)$ is positive unless $v_c ({\bm Q}^*)  = U + 2 V({\bm Q}^*) < 0$, 
so that $\tilde{f}_0^{a, {\rm AL}}$ increases. 
This observation of $\tilde{f}_0^{a, {\rm AL}}$ based on the EHM leads us to a general trend that  
charge fluctuations make $\chi_s$ increase, while spin fluctuations make $\chi_s$ decrease.

\subsection{Comparison with experiments}

The result obtained in this paper may be compared qualitatively with experiments on the spin susceptibility
of quasi-two-dimensional organic materials showing CO. Most relevant materials
are $\theta$-(BEDT-TTF)$_2$ RbZn$_4$ and $\beta$-({\it meso}-DMBEDT-TTF)$_6$PF$_6$.
As mentioned in \S~1, the spin susceptibility of the former material {\it increases} as temperature
is lowered from the room temperature to the metal-insulator transition temperature $T_{\rm MI} = 195{\rm K}$;
its value changes gradually from $\chi_s(T=300{\rm K}) = 6\times 10^{-4} {\rm emu \ mol}^{-1}$ 
to $\chi_s(T\simeq T_{\rm MI}) = 7\times 10^{-4}{\rm emu \ mol}^{-1}$\cite{H.Mori1}. 
Similarly, the spin susceptibility of the latter material {\it increases} from 
$\chi_s(T=300{\rm K}) = 6\times 10^{-4} {\rm emu \ mol}^{-1}$
to $\chi_s(T=T_{\rm MI} = 90{\rm K}) = 1.1\times 10^{-3} {\rm emu \ mol}^{-1}$ with decreasing
temperature\cite{H.Mori2}. These behaviors in the metallic phase are consistent with our theoretical result.
For more quantitative comparison, however, we need to consider several effects neglected
in the present calculation such as electron-phonon interaction, inclusion of realistic band dispersion,
and long-range Coulomb interaction beyond nearest-neighbor molecules.
In particular, we have to note that the present calculation is based on the weak-coupling approach
starting with a noninteracting metallic state; strong electron correlation, which
cannot be treated in the present approach, may affect the behavior of the spin susceptibility.
Nevertheless, the present calculation gives one of
possible explanations to enhancement of the spin susceptibility in terms of charge fluctuations developed 
toward the CO transition. 

It is comprehensive to consider the spin susceptibility of $\kappa$-(ET)$_2 X$ and $\beta$-(ET)$_2$I$_3$,
which are composed of two-dimensional conducting ET molecules forming dimers\cite{Kanoda06}.
These materials may be described by the Hubbard model at the half-filling (one hole per dimer) due to
molecular dimerization. Therefore, they show a metal-insulator transition by band-width control
with applying pressure or replacing anions. Antiferromagnetic order appears for
low temperatures in the insulating side, while superconductivity shows up in the metallic side.
A most metallic material $\beta$-(ET)$_2$I$_3$ shows a nearly temperature-independent
spin susceptibility ($\chi_s \sim 4 \times 10^{-4} {\rm emu \ mol}^{-1}$), i.e., 
the Pauli paramagnetism from the room temperature down to
the superconducting transition at $T = 3.6 {\rm K}$. $\kappa$-(ET)$_2$Cu(NCS)$_2$ and 
$\kappa$-(ET)$_2$Cu[N(CN)$_2$]Br with a narrower band width are closer to the antiferromagnetic
phase in the phase diagram, but still in the metallic side. The spin susceptibility of these two materials
takes a similar value as $\beta$-(ET)$_2$I$_3$ for high temperatures, but it {\it decreases}
as temperature is lowered. This experimental observation is 
consistent with our expectation that the increase of spin fluctuations make the spin susceptibility {\it decrease}
as discussed in the previous subsection. 

One advantage of our approach is that two opposite effects on the spin susceptibility by development of
charge  and spin fluctuations can be understood naturally in the weak-coupling approach.
We conjecture that this result is valid for many organic materials 
with a simple band structure
unless higher-order vertex corrections
become relevant just near the transition point.

Finally, we shall give brief comments on other CO materials. (DI-DCNQI)$_2$-M (M$=$Li and Ag) and
(TMTTF)$_2 X$ ($X=$monovalent anion) are famous quasi-one-dimensional conductors showing
CO\cite{H.Seo1, T.Takahashi}, but are out of the scope of the present study because
low-energy excitations in these systems may be described by the Tomonaga-Luttinger liquid 
in which spin and charge degrees of freedom are separated. $\alpha$-(BEDT-TTF)$_2$I$_3$ is also a
well-known CO organic conductor with a quasi-two-dimensional crystal structure~\cite{T.Takahashi}.
The spin susceptibility of this material decreases as temperature is lowered toward the CO
transition point at $T = 135{\rm K}$\cite{Rothaemel86}. We conjecture that this behavior apparently opposite to
our result is related to its multi-band structure with four small Fermi surfaces~\cite{Tajima06}. 
The present theoretical discussions on the single-band EHM can be generalized straightforwardly to a 
multi-band model by replacing the interaction with an interaction matrix. 
This extension is, however, non-trivial to be solved in a future problem.
The CO phenomena have also been observed in several inorganic materials.
For example, the vanadium bronze $\beta$-Na$_{0.33}$V$_2$O$_5$ shows the metal-insulator transition
at $T = 135~{\rm K}$ accompanied by CO~\cite{Yamada99,Itoh00,Suzuki06}. 
The spin susceptibility of this material is enhanced toward the metal-insulator transition with decreasing temperature
as indicated by our theoretical result for a simple one-band model, although our weak-coupling approach may be 
insufficient for this material with relatively large Coulomb interaction.

 
\section{Summary}
In this paper, we developed the non-skeleton diagrammatic expansion to satisfy the compressibility and 
spin-susceptibility sum rules. Based on this expansion, the leading vertex corrections to the static response 
functions have been examined for the two-dimensional extended Hubbard model in the vicinity of its 
charge-ordering transition point. 
It was shown that the charge-ordering transition is suppressed by charge fluctuations with the overall shape of 
its transition curve almost unchanged. 
We found that the same charge fluctuations make the uniform spin susceptibility {\it increase} due to interesting 
coupling of the charge and spin degrees of freedom through the vertex corrections. 

The increase in the spin susceptibility originates from the Maki-Tompson and Aslamazov-Larkin type 
vertex corrections. It is, however, noted that the former contribution is almost canceled by the self-energy 
correction, so that the net enhancement of the spin susceptibility can be controlled by the latter contribution. 
A physical interpretation of this result was also given in the Fermi-liquid theory: 
When charge fluctuations develop toward the transition, the effective mass of the quasi-particle tends to decrease 
but an attractive part in the spin channel of the quasi-particle interaction tends to increase. 
Since the latter effect overcomes the former, the uniform spin susceptibility gets enhanced.      

Our result of the enhanced spin susceptibility towards the charge-ordering transition 
is consistent with experiments on  $\theta$-(BEDT-TTF)$_2$RbZn$_4$\cite{H.Mori1} and 
$\beta$-({\it meso}-DMBEDT-TTF)$_2$PF$_6$\cite{H.Mori2}.

 
\section*{Acknowledgment}
K.~Y. thanks to H.~Mori and A.~Kobayashi for useful discussions.
H.~M. thanks to Y.~Takada for valuable discussions on conserving approximations.
T.~K thanks to Y.~Ueda for discussion on experiments of vanadium bronzes.
This work was supported by a Grant-in-Aid for Scientific Research in Priority Area of Molecular Conductors (No. 1828018) from the Ministry of Education, Culture, Sports, Science and Technology.
The computation in this work has partly been done by the facilities of the Supercomputer Center, 
Institute for Solid State Physics, University of Tokyo.


\appendix
\section{Systematic inclusion of vertex corrections}

In this Appendix, we present an iterative approximation scheme for systematic inclusion of 
vertex corrections to the charge and spin response functions.
We construct (A) the single-particle Green's function $G_{\sigma}^{(i)}(k)$, (B) the vertex function 
$\Lambda_{\sigma\sigma'}^{(i)}(k;q)$, and (C) the response functions $\chi_{_{NN}}^{(i)}(q)$ 
and $\chi_{_{S_z S_z}}^{(i)}(q)$. These functions are constructed so that the Ward identity holds
at each level of approximation assigned by an integer $i$,
and the next level of approximation assigned by $i+1$ is constructed from the $i$th approximation.
Our algorithm of generating a series of approximations is almost the same as the one
proposed by Takada~\cite{Takada95}. Our scheme differs from Takada's one on three points;
(1) we start with the mean-field Green's function as the initial ($i = 0$) approximation,
(2) we use the non-skeleton diagrammatic analysis for $\tilde{\Sigma}^{(i)}_{\sigma}(k)$
with respect to the {\it combination} of dressed Green's function, $G^{(i-1)}_{\sigma}(k)$ and $G^{(i)}_{\sigma}(k)$, 
and (3) we simplify iterative procedure by avoiding direct integral for the Bethe-Salpeter equations.
Our scheme is suitable to study the leading contribution of the vertex corrections.

\begin{figure}[t]
\begin{center}
\resizebox{70mm}{!}{\includegraphics{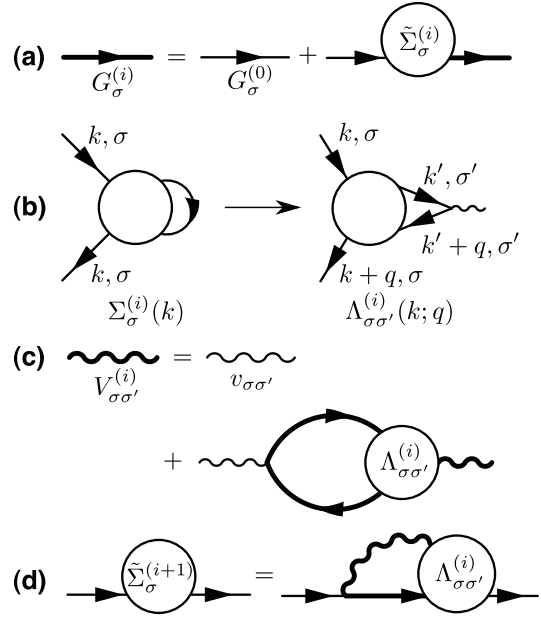}}
\end{center}
\caption{The Feynman diagrams of the present iterative approximation:
(a) the Dyson equation, (b) construction of the vertex function, (c) the effective
interaction, and (d) the next definition of the self-energy.}
\label{fig:diagram3}
\end{figure}

(A) {\it The single-particle Green's function}: 
Suppose the self-energy $\tilde{\Sigma}_{\sigma}^{(i)}(k)$ in which the Hartree contribution is subtracted 
is given as a functional of the 
mean-field Green's functions $G^{(0)}_{\sigma}$ ($\sigma = \pm$), i.e. 
\begin{align}
\tilde{\Sigma}_{\sigma}^{(i)}(k) \equiv \tilde{\Sigma}_{\sigma}^{(i)}(k: [G^{(0)}_{\pm}])\, .
\end{align} 
We take the initial function as $\tilde{\Sigma}_{\sigma}^{(0)}(k)$ $=$ $0$
(the Hartree approximation). Let the single-particle Green's function be 
\begin{align}
G_{\sigma}^{(i)}(k) =[ G_{\sigma}^{(0)}(k)^{-1} - \tilde{\Sigma}_{\sigma}^{(i)}(k) ]^{-1}\, ,
\end{align}
where the corresponding diagram is shown in Fig.~\ref{fig:diagram3}~(a).
$G_{\sigma}^{(i)}(k)$ is a function of $\tilde{\mu}_{\sigma}^{(i)}$; the particle number
and chemical potential are calculated as
\begin{align}
n_{\sigma}^{(i)} &= \int_{k}  {\rm e}^{{\rm i}\epsilon \eta}G_{\sigma}^{(i)}(k)\,, \\
\mu_{\sigma}^{(i)} &= \tilde{\mu}_{\sigma}^{(i)}+ 
\sum_{\sigma'}v_{\sigma\sigma'}({\bm 0}) n_{\sigma'}^{(i)} \, .
\label{density ITR}
\end{align}
In this paper, we fix $\tilde{\mu}_{\sigma}^{(i)} \equiv \tilde{\mu}_{\sigma}$ during the calculation,
and determine $\tilde{\mu}_\sigma$ by $n_{\sigma}^{(i)} = 3/4$ from a $\tilde{\mu}_{\sigma}$-$n$ curve.

(B) {\it The vertex function}: Following the Kadanoff-Baym scheme, the vertex function
$\Lambda_{\sigma\sigma'}^{(i)}(k;q)$ $\equiv$ $\Lambda_{\sigma\sigma'}^{(i)}(k;q: [G^{(0)}_{\pm}])$
is given as the sum of $\delta_{\sigma \sigma'}$ and all the possible diagrams obtained by
inserting the external vertex carrying a momentum $q$ into an arbitrary $G^{(0)}$-line with a 
spin $\sigma'$ in each diagram for $\tilde{\Sigma}_{\sigma}^{(i)}(k: [G^{(0)}_{\pm}])$. 
This procedure is schematically shown by the Feynman diagrams 
in Fig.~\ref{fig:diagram3}~(b). Note that the wave numbers
and the Matsubara frequencies at all the internal and external vertexes are conserved. 
The vertex function thus obtained satisfies various types of the Ward identity, 
one of which is given by
\begin{align}
\Lambda_{\sigma \sigma'}^{(i)}(k; 0) 
&= \delta_{\sigma \sigma'} - 
\frac{\partial \tilde{\Sigma}_{\sigma}^{(i)}(k)}{\partial \tilde{\mu}_{\sigma'}}\, .
\label{the i th Ward_q-limit}
\end{align}
This identity is proved by the same way as in \S~\ref{sec:2-3} as follows:
Since $\tilde{\Sigma}_{\sigma}^{(i)}(k)$ depends on $\tilde{\mu}_{\pm}$ only through 
$G^{(0)}_{\pm}$, the differential operation on $\tilde{\Sigma}_{\sigma}^{(i)}(k)$ with 
respect to $\tilde{\mu}_{\sigma}$ corresponds
to an operation to pick up one internal $G^{(0)}_{\sigma'}(p)$ line
in $\tilde{\Sigma}_{\sigma}^{(i)}(k)$ in all the possible ways, and to replaces it 
with $G^{(0)}_{\sigma'}(p)^2$. 
This differential operation is nothing but the vertex insertion for $q=0$
as seen in Fig.~\ref{fig:diagram3}~(b).

(C) {\it The response functions}: The response function $\chi_{\sigma \sigma'}^{(i)}(q)$
is written in terms of the one-interaction irreducible part $\tilde{\chi}_{\sigma \sigma'}^{(i)}(q)$ 
at the $i$th level of approximation as
\begin{subequations} 
\begin{align}
\chi_{\sigma \sigma'}^{(i)}(q)  &= 
\tilde{\chi}_{\sigma \sigma'}^{(i)}(q) - 
\sum_{\sigma_1, \sigma_2} \tilde{\chi}_{\sigma \sigma_1}^{(i)}(q) v_{\sigma_1 \sigma_2}({\bm q}) 
\chi_{\sigma_2 \sigma'}^{(i)}(q)\,,
\label{partial response function ITR} 
\\
\tilde{\chi}_{\sigma \sigma'}^{(i)}(q)  &= - \int_{k} G_{\sigma}^{(i)}(k+q)G_{\sigma}^{(i)}(k) 
\Lambda_{\sigma \sigma'}^{(i)}(k; q)\,.
\label{vertex corrections ITR}
\end{align}
\end{subequations}
Then, the charge and spin response functions $\chi_{_{NN}}^{(i)}(q)$ and $\chi_{_{S_z S_z}}^{(i)}(q)$ 
are calculated as
\begin{subequations}
\label{def: response functions ITR}
\begin{align}
\chi_{_{NN}}^{(i)}(q)&= \frac{1}{2} \sum_{\sigma, \sigma'} \chi_{\sigma \sigma'}^{(i)}(q),\\
\chi_{_{S_z S_z}}^{(i)}(q)&= \frac{1}{2} \sum_{\sigma, \sigma'} \sigma \sigma' \chi_{\sigma \sigma'}^{(i)}(q).
\end{align}
\end{subequations}
As proved in \S~\ref{sec:2-1},
the Ward identity (\ref{the i th Ward_q-limit}) leads to the compressibility and spin-susceptibility 
sum rules at the $i$th level of approximation as follows:
\begin{subequations}
\label{chi_identity ITR}
\begin{align}
\chi_c^{(i)} &\equiv 
\frac{1}{2} 
\frac{\partial n^{(i)}}{\partial \mu}
= \chi_{_{NN}}^{(i)}(0), 
\label{chi_charge ITR}
\\
\chi_s^{(i)} &\equiv
\frac{1}{2}
\frac{\partial m^{(i)}}{\partial h}
= \chi_{_{S_z S_z}}^{(i)}(0). 
\label{chi_spin ITR}
\end{align}
\end{subequations}

(D) {\it The self-energy in the next level of approximation}: For a systematic improvement in
the approximation, we require that the vertex functions should be taken into account in a
manner consistent with the single-particle Green's function. This requirement makes us
to select the self-energy in the next level of approximation as
\begin{subequations}
\begin{align}
\tilde{\Sigma}_{\sigma}^{(i+1)}(k) = 
- \sum_{\sigma'} \int_{q} G_{\sigma}^{(i)} (k+q) V_{\sigma \sigma'}^{(i)}(q) 
\Lambda_{\sigma \sigma'}^{(i)}(k;q)\, ,
\label{excess self-energy ITR}
\end{align} 
where the effective interaction $V_{\sigma \sigma'}^{(i)}(q)$ is given by
\begin{align}
V^{(i)}_{\sigma \sigma'}(q)  = v_{\sigma \sigma'}({\bm q}) 
- \sum_{\sigma_1, \sigma_2}v_{\sigma\sigma_1}({\bm q})\tilde{\chi}_{\sigma_1 \sigma_2}^{(i)}(q) 
V^{(i)}_{\sigma_2 \sigma'}({\bm q})\, .
\end{align}
\end{subequations}
The diagrams of these equations are shown in Fig.~\ref{fig:diagram3}~(c) and (d).
With $\tilde{\Sigma}_{\sigma}^{(i+1)}(k) \equiv \tilde{\Sigma}_{\sigma}^{(i+1)}(k: [G^{(0)}_{\pm}])$, 
we can construct the approximation for $\chi_{_{NN}}^{(i+1)}(q)$ and $\chi_{_{S_z S_z}}^{(i+1)}(q)$ 
following the processes in (A), (B) and (C).

In the text, we have used two approximation procedures described in \S~2.2 and in \S~2.3. 
In the present iterative approximation scheme, 
the former corresponds to $i=0$, and the latter to $i=1$. In principle, we can continue
the iterative process as one hopes, although it is difficult to continue it for $i\ge 2$ practically.

We note that the approximate response functions approach the exact ones
in the limit of $i \rightarrow \infty$ as 
\begin{subequations}
\label{def: response functions ITR}
\begin{align}
\lim_{i \rightarrow \infty} \chi_{_{NN}}^{(i)}(q) &=  \chi_{_{NN}} (q) \, , \\
\lim_{i \rightarrow \infty} \chi_{_{S_z S_z}}^{(i)}(q) &= \chi_{_{S_zS_z}} (q) \, .
\end{align}
\end{subequations}


\begin{thebibliography}{11}
\bibitem{footnote2}
In the RPA, another reentrant transition (metal $\rightarrow$ CO with further decreasing temperatures) occurs
as shown by a small return of the transition curve for $T \lesssim 0.1$ in Fig.~\ref{phase_U3.eps}.
This reentrant transition seems to be suppressed in the 1NSCA, although one needs
carefull calculations for this low-temperature region to clarify it.
\bibitem{footnote4}
$\tilde{\chi}^{\rm SC}_{V}$ is evaluated by taking the Fock-type diagram into account for the self-energy.
\end{thebibliography}
\end{document}